\documentclass{JHEP3}
\usepackage[dvips]{graphicx}  

\def\be{\begin{eqnarray}}
\def\ee{\end{eqnarray}}
\def\bea{\begin{eqnarray}}
\def\eea{\end{eqnarray}}

\newcommand{\eq}[1]{Eq.~(\ref{eq:#1})}
\newcommand{\sect}[1]{Sec.~\ref{sec:#1}}
\newcommand{\fig}[1]{Fig.~\ref{fig:#1}}

\newcommand{\she}{\sinh^2{\eta}}
\newcommand{\che}{\cosh^2{\eta}}
\newcommand{\lsads}{L_{\rm SAdS}}
\newcommand{\ls}[1]{L_s^{\rm #1}}

\newcommand{\tily}{\tilde{y}}

\newcommand{\calF}{{\cal F}}

\newcommand{\largeeta}{\xrightarrow{\eta \gg 1}}

\newcommand{\Ks}{K_*}
\newcommand{\sx}{\sigma_x}
\newcommand{\sr}{\sigma_r}
\newcommand{\sgh}{\sigma_h}
\newcommand{\deltakt}{\delta_{\rm cascade}}
\newcommand{\deltadp}{\delta_{Dp}}

\newcommand{\RNAdS}{RN-${\rm AdS}_5\:$}
\newcommand{\SAdS}{${\rm SAdS}_5\:$}
\newcommand{\SAdSd}{${\rm SAdS}_{d+1}\:$}

\usepackage{amsmath,amsfonts,amssymb}
\numberwithin{equation}{section}
\newcommand{\calE}{\mathcal{E}}

\newcommand{\hp}{\mathfrak{p}}
\newcommand{\hq}{\mathfrak{q}}
\newcommand{\tilp}{\Tilde{p}}
\newcommand{\tilq}{\Tilde{q}}
\newcommand{\tilI}{\Tilde{I}}

\newcommand{\barq}{\bar{q}}

\title{Screening length in plasma winds}

\author{Elena C\'aceres \\
Facultad de Ciencias,\\ Universidad de Colima,\\
Bernal D\'{\i}az del Castillo 340, Colima, Colima, M\'exico\\
\email{elenac@ucol.mx}}

 \author{Makoto Natsuume\\Theory Division, Institute of Particle 
and Nuclear Studies, \\
KEK, High Energy Accelerator Research
Organization,\\ Tsukuba, Ibaraki, 305-0801, Japan\\
\email{makoto.natsuume@kek.jp}}

\author{Takashi Okamura\\
Department of Physics,\\   Kwansei Gakuin University\\
Sanda, 669-1337, Japan\\
\email{okamura@ksc.kwansei.ac.jp}}

\preprint{UTTG-11-06, KEK-TH-1095} 
\abstract{
We study the screening length $L_s$ of a heavy quark-antiquark pair in strongly
coupled gauge theory plasmas flowing at velocity $v$. Using the AdS/CFT
correspondence we investigate, analytically, the screening length in the
ultra-relativistic limit. We develop a procedure that allows us to find the
scaling exponent for a large class of backgrounds.  We find that  for conformal
theories the screening length is $(\mbox{boosted energy density})^{-1/d}$.  As
examples of conformal backgrounds we study   R-charged black holes and
Schwarzschild-anti-deSitter black holes in $(d+1)$-dimensions.  For
non-conformal theories, we find that  the exponent  deviates from $-1/d$.  
As
examples we study the  non-extremal Klebanov-Tseytlin  and D$p$-brane
geometries. We find an interesting relation between the deviation of the
scaling exponent from the conformal value and the speed of sound.
}


\begin{document}

\section{Introduction}

The analysis of the quark gluon plasma (QGP) produced in relativistic heavy ion collisions  is, undoubtedly, a challenging task. One of the main difficulties is that in the temperature range of  current and near-future experiments QCD is believed to be strongly coupled. Thus, one often focuses on the generic signatures of QGP. For example, 
\begin{enumerate}
\item The elliptic flow
\item The jet quenching 
\item $J/\Psi$-suppression.
\end{enumerate}

The first signature, the elliptic flow, is interpreted as a consequence of the QGP having a very low viscosity \cite{Teaney:2003pb,Hirano:2005wx}. In \cite{Policastro:2001yc}, Policastro, Son and Starinets  pioneered the use of the AdS/CFT correspondence to study a strongly coupled plasma. It was found that the AdS/CFT prediction of the shear viscosity is in very good agreement  with the value derived from the elliptic flow \cite{Kovtun:2003wp}-\cite{Maeda:2006by}. (See Ref.~\cite{Natsuume:2007qq} for a review.) This success indicates that AdS/CFT might be a good tool to gain some insight in the physics of  QGP.  It is thus  natural to ask what are the AdS/CFT descriptions of the other  QGP signatures and what can we learn from them. In fact,  AdS/CFT descriptions  of jet quenching  and energy loss have  been proposed recently \cite{Liu:2006ug}-\cite{Gubser:2006bz}. (See Refs.~\cite{Buchel:2006bv}-\cite{Armesto:2006zv} for extensions to other  backgrounds and Ref.~\cite{Caceres:2006as} for comparison of these different proposals. See also Ref.~\cite{Sin:2004yx} for an early attempt.)

The next natural step is to investigate  $J/\Psi$-suppression. Since $J/\Psi$ is
heavy, charm pair production occurs only at the early stages of the nuclear
collision. However, if the production occurs in the plasma medium, charmonium
formation is suppressed due to Debye screening. One technical difficulty  is
that the $c\bar{c}$ pair is not produced at rest relative to the plasma.
Therefore, it is expected that the screening length will be  velocity dependent.
A dynamical calculation of the screening potential has been  done only for the
Abelian plasma \cite{Chu:1988wh}.

Recently, there has been an  interesting proposal by Liu, Rajagopal and
Wiedemann to model charmonium suppression via the AdS/CFT  correspondence
\cite{Liu:2006nn}.  The authors  considered  a boosted black hole and computed
the screening length in the quark-antiquark rest frame. In 
Ref.~\cite{Chernicoff:2006hi}, the authors studied the equivalent problem of a
quark-antiquark pair moving in a thermal background and calculated  the
screening length as well.
The main lessons drawn from Ref.~\cite{Liu:2006nn} are,
\begin{enumerate}
\item[(i)] The screening length is proportional to $(\mbox{boosted energy density})^{-1/4}$.
\item[(ii)] Aside from the boost factor, $(1-v^2)^{1/4}$, \ the screening length has only a mild dependence on the wind velocity $v$.
\item[(iii)] The length is minimum when $\theta=\pi/2$ and maximum when $\theta=0$, where $\theta$ is the angle of the quark-antiquark pair (dipole) relative to the plasma wind.
\end{enumerate}
The main focus in Ref.~\cite{Liu:2006nn} is the five-dimensional Schwarzschild-AdS black hole (${\rm SAdS}_5$), which is dual to the ${\cal N}=4$ super-Yang-Mills theory (SYM) at finite temperature. Even in that simple background,  numerical computations were needed in order to see the full details of the screening length.

In this paper, we focus on the ultra-relativistic limit, where analytic
computations are possible. This makes it easier to carry out the analysis in
various, more involved, backgrounds. 
We study the scaling of the screening length with the energy density and the velocity. Our results are summarized as follows: 
\begin{enumerate}
\item 
For conformal theories, we find the behavior 
\be
(\mbox{screening length}) \propto (\mbox{boosted energy density})^{-1/d}~,
\label{eq:result1}
\ee
where $d$ denotes the number of dimensions of a dual gauge theory. Examples are \SAdSd and R-charged black holes with three generic charges. 

\item In particular, in the ultra-relativistic limit, the screening length at
finite chemical potential is the same as the one at zero chemical potential {\it for a given energy density}.

\item
For non-conformal theories,%
\footnote{
In this paper, we use the word ``non-conformal" for a theory with nontrivial dilaton.}
the exponent deviates from $1/d$. Examples are the non-extremal
Klebanov-Tseytlin (KT) geometry and the D$p$-brane solution. 
For small nonconformality, the deviation is proportional to the parameter of the nonconformality. We conjecture a universal expression of the deviation written in terms of the speed of sound.

\end{enumerate}
These results are in a sense natural;  conformal theories have only 
few dimensionful parameters so the screening length should behave as
\eq{result1} from dimensional grounds. On the other hand, non-conformal
theories have  other dimensionful parameters, so the screening length is not
determined from dimensional analysis. However, even for conformal theories, 
the details are not determined from dimensional analysis alone. For example, in
the ultra-relativistic limit, the screening length in a R-charged black hole
background
 is independent of the chemical potential.
Also, for non-conformal theories, the exponent turns out to be {\it smaller}
than the
one for the conformal examples.

In the next section, we will derive the relevant equations of motion  for
general backgrounds. We then proceed to analyze conformal theories in
\sect{conf}, where we also develop a general procedure to treat a large class of
backgrounds. As examples, we calculate the screening length for R-charged black
holes and ${\rm SAdS}_{d+1}$.  In \sect{nonconf}, we analyze two non-conformal
theories: the non-extremal Klebanov-Tseytlin solution  and D$p$-branes
backgrounds. 
We conclude in \sect{discussion} with discussion, implications and future
directions.

\section{General setup}

In the AdS/CFT framework, a heavy quark may be realized by a fundamental string which stretches from the asymptotic infinity (or from a ``flavor brane") to the black hole horizon. This string transforms as a fundamental representation; In this sense, the string represents a ``quark." The fundamental string has an extension and the tension, so the string has a large mass, which means that the string represents a heavy quark. 

For a  $q\bar{q}$ pair, two individual strings extending to the boundary is not
the lowest energy configuration. Instead, it is energetically favorable to have
a single string that connects the pair. The energy difference is interpreted as
a $q\bar{q}$ potential and  has been widely studied in the past from the AdS/CFT
perspective. At finite temperature \cite{Rey:1998bq,Brandhuber:1998bs}, it is no
longer true that a string connecting the $q\barq$ pair is always  the lowest
energy configuration; for large enough separation of the pair, isolated strings
are favorable energetically. This phenomenon is the dual description of Debye
screening in  AdS/CFT. In Ref.~\cite{Liu:2006nn}, the authors computed
$q\bar{q}$ screening length in the $q\bar{q}$ rest frame, {\it i.e.}, they considered
the plasma flowing at a velocity $v$. Such a ``plasma wind" is obtained by
boosting a black hole background.

\subsection{Equations of motion}\label{sec:eom}

\begin{figure}[tb]
\begin{center}
\includegraphics{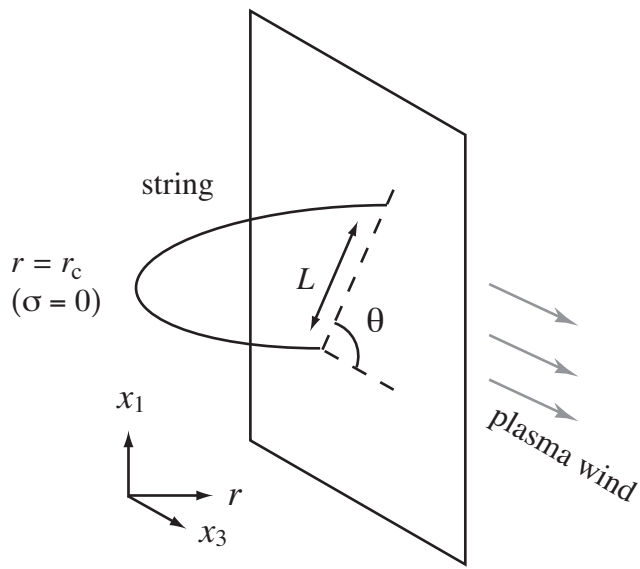}
\vskip2mm
\caption{The fundamental string connecting the quark-antiquark pair. (The shape of the string should not be taken seriously.)}
\label{fig:wind}
\end{center}
\end{figure}

In order to consider the screening length in the dipole rest frame, we boost the  background metric. We assume an unboosted metric of the form
\be
ds^2 = g_{xx}\,\{-(1-h)dt^2+dx_i^2\}+g_{rr}\,dr^2 +\cdots~.
\ee
Consider the plasma wind in the $x_3$-direction, so the boosted metric is
\bea
\lefteqn{-(1-h) dt^2+dx_3^2}
\nonumber\\
&& \rightarrow -(1-h \che)dt^2-2h \sinh\eta \cosh\eta\, dt dx_3+(1+h \she)dx_3^2~,
\eea
where $v$ is the wind velocity, $\cosh\eta = \gamma$, \ \ $\sinh\eta = \gamma v$, and $\gamma=1/\sqrt{1-v^2}$.

The dynamics of the fundamental string is governed by the Nambu-Goto action. The quark-antiquark pair is chosen to lie in the $(x_1,x_3)$-plane at an angle $\theta$ relative to the wind (See \fig{wind}). Thus, we choose the gauge $\tau = t,~\sigma = x_1$ and consider the configuration
\be
  x_3 = x_3(\sigma)~, \quad 
  r = r(\sigma)~, \quad 
  \mbox{constant otherwise}.
\label{eq:configuration}
\ee
Then, the Nambu-Goto action (in the string metric%
\footnote{In this paper, we mostly use the string metric. If one would like to work in the $d=10$ Einstein metric, simple make the following replacements in all formulae in this subsection: 
$g_{xx} \rightarrow e^{\phi/2}g_{xx},
g_{rr} \rightarrow e^{\phi/2}g_{rr}$.
}
) becomes
\bea
S = \frac{-1}{2\pi l_s^2}\int d\sigma^2~ {\cal L} 
&=& \frac{-1}{2\pi l_s^2}\int d\sigma^2~ 
\sqrt{- \mbox{det}\, G_{ab}} \\
&=& \frac{-1}{2\pi l_s^2}\int d\sigma^2~ 
\sqrt{|G_{00}|(G_{11}+G_{33}x_3'^2+G_{rr}r'^2)+G_{03}^2x_3'^2}~.
\eea
Here, we used the boosted metric components such as $G_{03}$ and $~{}' = d/d\sigma$. The conserved quantities are
\bea
- q &:=& \left(\frac{\partial {\cal L}}{\partial r'} \right) r'
+ \left(\frac{\partial {\cal L}}{\partial x_3'}\right) x_3'-{\cal L}~, \\
p &:=& \frac{\partial {\cal L}}{\partial x_3'}~,
\eea
which become
\bea
q^2 r'^2 &=& \frac{g_{xx}}{g_{rr}} \left[
\frac{1-h \che}{1-h}\left\{ 
g_{xx}^2(1-h)-p^2 \right\} - q^2 \right]
=: \calF (r)~,
\label{eq:EOM_r} \\
q~x_3' &=& p\,\frac{1-h \che}{1-h}~,
\label{eq:EOM_x3}
\eea
or equivalently,
\bea
x_1 &=& \sigma = q \int \frac{dr}{\sqrt{\calF}}~,
\label{eq:eom1} \\
x_3 &=& \int dr \frac{dx_3}{dr} 
= p \int \frac{dr}{\sqrt{\calF}}\frac{1-h\che}{1-h} \nonumber\\
&=& p \left[ \frac{x_1}{q} - \she \int \frac{dr}{\sqrt{\calF}}\frac{h}{1-h}\right]~.
\label{eq:eom3}
\eea

The string stretches from  asymptotic infinity and reaches a turning point
$r=r_c$ defined by $\calF(r_c)=0$. Then, the string goes back to  asymptotic
infinity. From the symmetry, the turning point occurs at $\sigma=0$.  Then the
 boundary conditions are summarized as, 
\be
\begin{array}{ll}
  r(\sigma=0) = r_c~, \quad & \calF(r_c) = 0~,
\\
  r(\sigma=L \sin\theta/2) = \infty~, \quad &
  x_3(\sigma=L \sin\theta/2) = \frac{L}{2}\, \cos\theta~.
\end{array}
\ee
where $L$ is the dipole length. The boundary conditions determine the
integration constants $p$ and $q$ in terms
of $L$ and $\theta$:
\bea
  \frac{L}{2}\, \sin\theta
  &=& q\, I_s(p, q, \eta)~,
\label{eq:L_sin} \\
  \frac{L}{2}\, \cos\theta
  &=& p\, \Big[~I_s(p, q, \eta)
  - \sinh^2\eta\, I_c(p, q, \eta)~\Big]~,
\label{eq:L_cos}
\eea
where
\bea
  I_s(p, q, \eta)
  &:=& \int^\infty_{r_c} \frac{dr}{ \sqrt{ \calF } }~,
\label{eq:def-I_s} \\
  I_c(p, q, \eta)
  &:=& \int^\infty_{r_c} \frac{dr}{ \sqrt{ \calF } }\,
                       \frac{h}{ 1 - h }~.
\label{eq:def-I_c}
\eea
The energy is given by
\bea
E &=& -\frac{S}{T}  \nonumber \\
&=& \frac{1}{\pi l_s^2} \int_{r_c}^\infty dr\, 
\sqrt{g_{xx}g_{rr}(1-h \che)}
\sqrt{ \frac{g_{xx}}{g_{rr}\calF} (q^2+p^2\frac{1-h \che}{1-h})+1 }~.
\eea
As usual, this energy can be made finite by subtracting the self-energy of 
a disconnected quark and antiquark pair which is 
\be
E_0 = \frac{1}{\pi l_s^2} \int_{r_0}^\infty dr\, 
\sqrt{g_{xx}g_{rr}(1-h \che)}~,
\ee 
where $r_0$ is the location of the horizon.

Two special cases are worth discussing separately.
\begin{enumerate}

\item[(i)] $\theta=\pi/2$:
This is the case when the plasma wind is perpendicular to the dipole. 
In this case, \eq{L_cos} implies $p=0$.
The integration constant $q$ can be determined from
\be
L 
= 2q \int _{r_c}^\infty 
\frac{ \sqrt{g_{rr}}~dr }
{ \sqrt{ g_{xx} \{(1-h \che)
g_{xx}^2-q^2 \} } }~.
\label{eq:perpendicular_formula}
\ee

\item[(ii)] $\theta=0$:
This is the case when the plasma wind is parallel to the dipole. 
In this case, \eq{L_sin} implies $q=0$. Then,
\be
L = 2p \int_{r_c}^\infty dr \frac{1}{\sqrt{\calF|_{q=0}}} \frac{1-h \che}{1-h}~.
\label{eq:parallel_formula}
\ee
\end{enumerate}

\subsection{AdS/CFT dictionary}\label{sec:dictionary}

Some well-known AdS/CFT dictionary is summarized here. 
This dictionary is valid for ${\rm AdS}_{5}$, R-charged black holes, and the D$p$-brane.
Roughly speaking, on the left-hand side we have  gravity quantities which are written in terms of gauge theory quantities on the right-hand side.
\bea
16\pi G_{10} &=& (2\pi)^7 g_s^2 l_s^8~, \\
R^{7-p} &=& \frac{(2\pi)^{7-p}}{(7-p) \Omega_{8-p}} g_s N_c l_s^{7-p}
\qquad \mbox{($R$ in terms of $N_c$)}, \\
2(2\pi)^{p-2}\, g_s\, l_s^{p-3} &=& g_{\rm YM}^2
\qquad \mbox{($g_s$ in terms of  $g_{\rm YM}$)},
\eea
where 
\be
\begin{array}{rl}
G_{10}:& \mbox{10-dimensional Newton constant}\\
g_s:& \mbox{string coupling} \\
R:& \mbox{AdS radius} \\
N_c:& \mbox{number of colors} \\
g_{\rm YM}:& \mbox{YM coupling} \\
\Omega:& S^n \mbox{ volume of unit radius; } 
\Omega_n=2\pi^{\frac{n+1}{2}}/\Gamma(\frac{n+1}{2})~.
\end{array}
\nonumber
\ee
When $p=3$, 
\bea
16\pi G_{10} &=& (2\pi)^7 \frac{R^8}{(4\pi N_c)^2}~, \\
(R/l_s)^4 &=& \lambda~,
\eea
where $\lambda$ is the 't~Hooft coupling. 
Below we often consider effective five-dimensional theories with the five-dimensional Newton constant given by 
\be
G_5 = \frac{G_{10}}{R^5 \Omega_5} = \frac{\pi R^3}{2N_c^2}~.
\ee

\section{Leading behavior of the screening length}\label{sec:leading}

\subsection{An example: \SAdS}\label{sec:sads}

As an example, let us start with the \SAdS case. This has been studied
previously in Refs.~\cite{Liu:2006nn,Chernicoff:2006hi}. Here we
study it 
analytically in the ultra-relativistic limit. The \SAdS black hole is given by
\be
  ds^2 = - \left(\frac{r}{R}\right)^2 
\left\{ 1-\left(\frac{r_0}{r}\right)^{4} \right\} dt^2 
+ \frac{dr^2}{\left(\frac{r}{R}\right)^2 \{ 1-(\frac{r_0}{r})^{4} \} } 
+ \left(\frac{r}{R}\right)^2 (dx_1^2+dx_2^2+dx_3^2)~.
\label{eq:SAdS5}
\ee
The temperature $T$ and the (unboosted) energy density $\epsilon_0$ of the black hole are given by 
\bea
T &=& \frac{r_0}{\pi R^2}, \\
\epsilon_0 &=& \frac{3}{16\pi G_5}\frac{r_0^4}{R^5}  = \frac{3}{8}\pi^2N_c^2T^4.
\eea

For $\theta=\pi/2$, \eq{perpendicular_formula} becomes
\be
\lsads(\theta=\pi/2) = 
2\hq\, 
\frac{R^2}{r_0}
\int_{y_c}^\infty dy \frac{1}{\sqrt{(y^4-1)(y^4-y_c^4)}}~,
\label{eq:perpendicular}
\ee
where we used the dimensionless variables
\be
y := \frac{r}{r_0}~, \quad
\hq := \left( \frac{R}{r_0} \right)^2 q~, \quad
\hp := \left( \frac{R}{r_0} \right)^2 p~,
\label{eq:rescaled_sads}
\ee
and
$y_c$ is the turning point given by $y_c^4:=\che+\hq^2$. 

The above integral \eq{perpendicular} can be evaluated numerically but is
slightly involved since one in interested in $\lsads$ as a function of $\hq$. We
choose not to follow the numerical route; Instead, we investigate analytically
the ultra-relativistic limit of \eq{perpendicular}.
For large $\eta$, $y_c \gg 1$, so the integral reduces to 
\bea
\lsads(\theta=\pi/2) &\largeeta& 
2\hq\, 
\frac{R^2}{r_0}
\int_{y_c}^\infty dy \frac{1}{y^2\sqrt{y^4-y_c^4}} \\
&=& 2\sqrt{\pi} \frac{\Gamma(3/4)}{\Gamma(1/4)} 
\frac{\hq}{(\che+\hq^2)^{\frac{3}{4}}} 
\frac{R^2}{r_0}~.
\eea

We are interested in the behavior of $\lsads$ as a function of $\hq$. The length $\lsads$ goes to zero both for small $\hq$ and for large $\hq$. Thus, there is a maximum value $L_s$ at some $\hq_m$. This means that there is no extremal world-sheet which binds the quark-antiquark pair for $L>L_s$, so this $L_s$ is defined as a screening length in Ref.~\cite{Liu:2006nn}.
The maximum of $\lsads$ is given by 
\bea
\ls{SAdS}(\theta=\pi/2) 
&=& \frac{2\sqrt{2\pi}}{3^{3/4}} \frac{\Gamma(3/4)}{\Gamma(1/4)}
\frac{1}{\sqrt{\cosh\eta}} 
\frac{R^2}{r_0} 
\qquad\mbox{at $\hq_m = \sqrt{2}\cosh\eta$}
\\
&\sim& \frac{0.743}{\sqrt{\cosh\eta}} 
\frac{R^2}{r_0}~.
\eea
Rewriting in terms of gauge theory variables, the screening length $L_s$ is given by 
\be
\ls{SAdS}(\theta=\pi/2) 
\sim \frac{0.743}{\sqrt{\cosh\eta}} 
\frac{1}{\pi T}~.
\label{eq:perpendicular_result0}
\ee
This parametrization in terms of temperature was  adopted in
\cite{Liu:2006nn,Chernicoff:2006hi}. One can equally express the result
in terms of energy density:
\bea
\ls{SAdS}(\theta=\pi/2) 
&=& \frac{2^{3/4}}{\sqrt{3}} \frac{\Gamma(3/4)}{\Gamma(1/4)} 
\frac{\sqrt{N_c}}{(\epsilon_0\che)^{1/4}} 
\label{eq:perpendicular_result} \\
&\sim& 0.328 \frac{\sqrt{N_c}}{(\epsilon_0\che)^{1/4}}~.
\eea
The choice of these parametrizations does not matter for \SAdS since there is no
other dimensionful quantities. But it does  matter when one has  other
dimensionful quantities such as the chemical potential. Our results in the next
section strongly indicate that one should really choose the energy density to
parametrize the screening length. However, one drawback of doing so is that  the
results will depend also on $N_c$ (and $\lambda$ for D$p$-branes), so
they are not finite in the $N_c \rightarrow\infty$ limit. If one chooses $T$, they do not appear
explicitly and appear only through $T$.

A few comments are in order. The temperature dependence in \eq{perpendicular_result0} is very natural and is the same as the standard Debye length. In terms of the energy density, $\sqrt{N_c}$ appears in the expression because $\epsilon_0 \propto N_c^2 T^4$ (both in the weak coupling and in the strong coupling as first noted by Gubser, Klebanov, and Peet~\cite{Gubser:1996de}.) On the other hand, the expressions have no dependence on the coupling and are in contrast to the naive weak coupling result $1/L_s \propto g_{\rm YM} \sqrt{N_c} T$. This is due to the $\lambda\rightarrow\infty$ limit; the coupling dependence should appear as subleading terms in the large-$\lambda$ expansion.

For $\theta=0$, \eq{parallel_formula} becomes
\be
\lsads(\theta=0) = 2\hp\, 
\frac{R^2}{r_0}
\int_{y_c}^\infty \frac{dy}{y^4-1}{\sqrt{\frac{y^4-\cosh^2\eta}{y^4-y_c^4}}}~,
\label{eq:parallel}
\ee
where $y_c^4:=1+\hp^2$. In order for the function inside the square-root not to become negative, $y_c>\sqrt{\cosh\eta}$. This suggests $\hp$ is large in this case. In terms of the rescaled variables 
$\hp =: \cosh\eta\,\tilp$ and
$y =: \sqrt{\cosh\eta}\,\tily$, 
the above integral in the large-$\eta$ limit becomes 
\be
\lsads(\theta=0) \largeeta
\frac{2\tilp}{\sqrt{\cosh\eta}} 
\frac{R^2}{r_0}
\int_{\sqrt{\tilp}}^\infty  \frac{d\tily}{\tily^4}{\sqrt{\frac{\tily^4-1}{\tily^4-\tilp^2}}}~.
\ee
The above integral can be evaluated numerically. The maximum of $\lsads$ is given by
\bea
\ls{SAdS}(\theta=0) 
&\sim& \frac{0.838}{\sqrt{\cosh\eta}}
\frac{R^2}{r_0}
\qquad\mbox{at $\hp_m = 1.38\cosh\eta$}
\label{eq:parallel_result} \\
&\sim& 0.370 \frac{\sqrt{N_c}}{(\epsilon_0\che)^{1/4}}~.
\eea
Hence, the screening length increases for the wind parallel to the dipole
compared with the screening length for the wind perpendicular to the dipole as
was observed numerically by Ref.~\cite{Liu:2006nn}. 

 Note also that doing a numerical fit to find the scaling at large velocities
can be misleading. In fact, in Ref.~\cite{Chernicoff:2006hi}, the authors found
 that the optimal  numerical fit valid for the entire  range $ 0 \le v \le 1$
 scales
like $(1-v^2)^{1/3}$. This could lead us to believe that $(1-v^2)^{1/3}$ is
also the correct  scaling for large $v$ which  we have seen  is not the
case. Analytically examining the behavior of $L_s$ in different
limits is a powerful tool;  We will apply it in the next sections to find the
scaling of $L_s$ as $v\rightarrow 1$.

\subsection{A formula for the scaling exponent}\label{sec:formula}

An analysis similar to the previous subsection is straightforward with the other
backgrounds. Instead of studying each background one by one, we will  develop  a
general formalism to find the scaling exponent. 

We start with Eqs.~(\ref{eq:L_sin})-(\ref{eq:def-I_c})
and $\calF$ defined by Eq.~(\ref{eq:EOM_r}),
\begin{align}
   \calF(r)
  &:= \frac{g_{xx}(r)}{g_{rr}(r)}~\left[~
    \frac{1 - h(r)\, \cosh^2\eta}{1 - h(r)}~
    \left\{ g_{xx}^2(r)\, \big( 1 - h(r) \big)
    - p^2 \right\} - q^2~\right]~.
\nonumber
\end{align}
The turning point $r_c$ is defined by $\calF(r_c) = 0$ and
satisfies
\begin{align}
   \cosh^2\eta
  &= \frac{1}{h(r_c)}~\left[~1 - q^2~
    \frac{ 1 - h(r_c) }{ g_{xx}^2(r_c) \big( 1 - h(r_c) \big) - p^2 }~
  \right]~.
\label{eq:cosh_eta}
\end{align}
%

We assume that 
the metric falls off
\begin{align}
  & g_{xx}(r)
  \sim \left( \frac{r}{R} \right)^{\sigma_x}~,
& & g_{rr}
  \sim C^2\, \left( \frac{r}{R} \right)^{-\sigma_r}~,
& & h(r) \sim \frac{m}{r^{\sigma_h}}
  = \frac{m}{R^{\sigma_h}}\, \left( \frac{r}{R} \right)^{-\sigma_h}~,
\label{eq:fall-off}
\end{align}
near the infinity $r = \infty$.
The parameter ``$m$" is the mass parameter. {\it A priori} there is no reason to regard it as the energy density, but in our examples below, it in fact represents the energy density. 
Furthermore, we assume that the metric behaves as%
\footnote{If one chooses the radial coordinate $r$ used for the ${\rm SAdS}_5$, this condition is equivalent to the condition $\sigma_h \ge 2\, \sigma_x > 0~$.}
%
\begin{align}
  & g_{xx}^2~h \sim \left(\text{at most~~} O(1)\right)~,
& & g_{xx} \sim (\text{divergent})~.
\label{eq:fall-off_index}
\end{align}


If $\sigma_x,~\sigma_h > 0$,
the turning point satisfies
$h(r_c) \ll 1$ for large-$\eta$ from Eq.~(\ref{eq:cosh_eta}),%
\footnote{The function,
$g(r) := g_{xx}^2(r) \left( 1 - h(r) \right) - p^2$, does not vanish.
Because the string has its end points at the infinity
and $g(\infty)=+\infty$,
the function $g(r)$ approaches to zero with positive value.
If $g(r)$ became positive infinitesimal value,
the RHS of Eq.~(\ref{eq:cosh_eta}) would be negative.
}
more precisely, $O\big( h(r_c) \big) = O\big( 1/\cosh^2\eta \big)$,
so that the turning point is near the infinity.
We are interested in the leading order term of $\cosh\eta$,
so we need only the leading term of the metric.

Using the parameter,
\begin{align}
  & \calE := \frac{m \cosh^2\eta}{R^{\sigma_h}}~,
\label{eq:def-calE}
\end{align}
the function $\calF$ behaves as
\begin{align}
  & \calF
  \largeeta \frac{1}{C^2}\, \left( \frac{r}{R} \right)^{\sigma_x + \sigma_r}
  \left[~\left( 1 - \calE\, \left( \frac{r}{R} \right)^{- \sigma_h}
         \right)
  \left\{ \left( \frac{r}{R} \right)^{2 \sigma_x} - p^2 \right\}
  - q^2~\right]~.
\label{eq:calF-approx}
\end{align}
Using the rescaled variables%
\footnote{For ${\rm SAdS}_5$, the coordinate $t$ is related to the coordinate $\tilde{y}$ (used in \sect{sads}) by $\tilde{y}^4=t$. The definition of conserved quantity $\tilp$ coincides with the one for ${\rm SAdS}_5$.}
\begin{align}
  & t := \frac{ (r/R)^{\sigma_h} }{\calE}~,
& & \tilp^2 := \frac{p^2}{\calE^{2 \sigma_x/\sigma_h}}~,
& & \tilq^2 := \frac{q^2}{\calE^{2 \sigma_x/\sigma_h}}~,
\label{eq:def-new_variables}
\end{align}
we can rewrite Eqs.~(\ref{eq:L_sin}) and (\ref{eq:L_cos}) as
\begin{align}
  & \frac{L}{R}\, \sin\theta
  \largeeta \frac{2\, C}{\sigma_h}~\calE^{-\nu}~\tilq~
  \tilI_s(\tilp, \tilq)~,
\label{eq:L_sin-approx} \\
  & \frac{L}{R}\, \cos\theta
  \largeeta \frac{2\, C}{\sigma_h}~\calE^{-\nu}~\tilp~
     \Big[~\tilI_s(\tilp, \tilq) - \tilI_c(\tilp, \tilq)~\Big]~,
\label{eq:L_cos-approx}
\end{align}
where
\begin{align}
  & \tilI_s(\tilp, \tilq)
  := \int^\infty_{t_c} dt~\frac{ t^{-\nu-1/2} }
    { \sqrt{ ( t - 1 ) \left( t^{2 \lambda}
            - \tilp^2 \right)- \tilq^2\, t } }~,
\label{eq:def-tilI_s} \\
  & \tilI_c(\tilp, \tilq)
  := \int^\infty_{t_c} \frac{dt}{t}~
    \frac{ t^{-\nu-1/2} }
         { \sqrt{ ( t - 1 ) \left( t^{2 \lambda}
                - \tilp^2 \right)- \tilq^2\, t } }~,
\label{eq:def-tilI_c}
\end{align}
and
\begin{align}
  & \nu := \frac{\sx+\sr-2}{2\sgh}~,
& & \lambda := \frac{\sigma_x}{\sigma_h} \le \frac{1}{2}~.
\label{eq:def-index}
\end{align}
The turning point $t_c \ge \text{max}(|\tilp|^{1/\lambda}, 1)$
is then determined by
\begin{align}
  & 0 =  ( t_c - 1 ) \left( t_c^{2 \lambda}
      - \tilp^2 \right)- \tilq^2\, t_c~.
\end{align}

Equations~(\ref{eq:L_sin-approx}) and (\ref{eq:L_cos-approx}) imply that 
the maximum of $L$, $L_s$, behaves as
\begin{align}
  & L_s
  \propto R \calE^{-\nu}
  \propto R \left( \frac{m}{R^{\sigma_h}}~\cosh^2\eta \right)^{-\nu}
\label{eq:L_max-behavior}
\end{align}
irrespective of $\theta$. 
Since the parameter $m$ is related to the energy density in our examples, the screening length $L_s$ is written in terms of
the boosted ``energy density" of plasma wind 
at large-$\eta$.%
\footnote{
One may wonder if $L_s$ is always written in terms of the combination $\epsilon\che$. This question however is not very meaningful when the other dimensionful quantities exist, so one should not take the combination very seriously. The exponent $\nu$ must be understood as the power of the (squared) Lorentz factor, $\che$.
}
Below we compute the exponent $\nu$ in various theories.

Finally, let us  briefly discuss the radial-coordinate dependence of our
formulae. We used the radial coordinate~$r$ for \SAdS and one may use a similar
coordinate which approaches  $r$ asymptotically when one considers more
general backgrounds. But it is sometimes more convenient to use a coordinate
other than $r$ (See, {\it e.g.}, the KT geometry in \sect{kt}). Thus, let us
check if the choice of a coordinate affects our discussion.
When we derive \eq{L_max-behavior}, we used only the power-law behavior of the metric (\ref{eq:fall-off}).
Also, the equations of motion, {\it e.g.}, Eqs.~(\ref{eq:eom1}) and
(\ref{eq:eom3}) themselves are invariant under the reparametrization of the
radial coordinate $r$. This is because the $g_{rr}$-component appears only in
the form of $\sqrt{g_{rr}}dr$. The power-law behavior together with the
reparametrization invariance suggests that one is free to choose any radial
coordinate as long as the metric has a
power-law behavior.
Indeed, consider a new coordinate $\bar{r}$ such as
\be
\frac{r}{R} = \left(\frac{\bar{r}}{R}\right)^\kappa \left( 1+O[(R/r)^a] \right), \qquad
a>0
\ee
and define new power indices $\bar{\sigma}_x, \bar{\sigma}_r$ and
$\bar{\sigma}_h$. The new indices still satisfy the
condition~(\ref{eq:fall-off_index}) even if $\kappa<0$. One can easily check
that the physical indices are all unchanged:
\be
\bar{\nu}=\nu~, \qquad
\bar{\lambda}=\lambda~, \qquad
\left| \frac{\bar{C}}{\bar{\sigma}_h} \right|
=\left| \frac{C}{\sgh} \right|~.
\ee
Therefore, our formulae are not affected by the choice or radial coordinate; 
One can choose a radial coordinate at will as long as the metric has a power-law
behavior.

\section{Conformal theories}\label{sec:conf}

For conformal theories, our results are consistent with the behavior
\be
(\mbox{screening length}) \propto (\mbox{boosted energy density})^{-1/d}~,
\ee
where $d$ denotes the number of dimensions of a dual gauge theory.

\subsection{R-charged black holes}

As a first example, consider five-dimensional black holes charged under the R-symmetry group $U(1)_R^3$. These backgrounds are dual to the ${\cal N}=4$ SYM with chemical potentials. The three-charge STU-solution (with noncompact horizon) is specified by the following background metric:
\footnote{
In this paper, we assume that the string satisfies the ansatz \eq{configuration}, namely the string is fixed in the compact dimensions when embedded into 10-dimensions. This means that the string we consider is not charged under the R-charges. When the ansatz fails, one should interpret the string as a ``smeared" string in the compact dimensions.
}
\begin{equation}
ds^2 = - {\cal H}^{-2/3}\, f \, dt^2 
+ {\cal H}^{1/3}  f^{-1} dr^2 
+ {\cal H}^{1/3} \left(\frac{r}{R}\right)^2 (dx_1^2 +dx_2^2 +dx_3^2)~,
\end{equation}
where
\begin{equation}
f = - {\mu\over r^2} + {r^2\over R^2}{\cal H} \,, \qquad 
H_i = 1 + {c_i\over r^2}\,, \qquad  {\cal H} = H_1 H_2 H_3 \,.
\end{equation}
The outer horizon $r_+$ is given by the larger root of $f(r)=0$.
The three R-charges $c_i$ are related to the  angular momenta $l_i$ in 
10-dimensions, $c_i=l_i^2$.

When three charges are equal, $c_i=c$, the STU-solution reduces to the Reissner-Nordstr\"{o}m-${\rm AdS}_5$ (RN-${\rm AdS}_5$) black hole. The standard form of the \RNAdS black hole is written as
\bea
  ds^2 &=& - f(r) dt^2 + f^{-1}(r) dr^2 + \left(\frac{r}{R}\right)^2 (dx_1^2+dx_2^2+dx_3^2)~,
\\
  f(r) &=& \left(\frac{r}{R}\right)^2 - \frac{m_{\rm RN}}{r^2} + \frac{q_{\rm RN}^2}{r^4}~.
\label{eq:RNAdS}
\eea
This form is related to the STU-solution by a coordinate transformation
$r^2+c \rightarrow r^2$ with $\mu=m_{\rm RN}$ and $\mu c=q_{\rm RN}^2$.

The temperature $T$ and the energy density $\epsilon_0$ of the black hole are given by 
\bea
T &=& 
{2 + \kappa_1 + \kappa_2 + \kappa_3 - \kappa_1 \kappa_2  \kappa_3\over 
2\sqrt{(1+\kappa_1)(1+\kappa_2) (1+\kappa_3)}}\, T_0\,.
\\
\epsilon_0 &=& 
\frac{3}{16\pi G_5}\frac{r_+^4}{R^5}\prod\limits_{i=1}^3(1+\kappa_i) 
= { 3 \pi^2 N^2 T_0^4 \over 8} \prod\limits_{i=1}^3 (1+\kappa_i)\,,
\label{eq:energy_density} 
\eea
where $T_0$ is the temperature without charges. Also, $\kappa_i$ is not the physical charge but the the rescaled charge $\kappa_i= c_i/r^2_+$.

The fall-off behavior of the STU-solution is
\be
g_{xx} \sim \left(\frac{r}{R}\right)^2, \quad
g_{rr} \sim \left(\frac{r}{R}\right)^{-2}, \quad
h \sim \prod\limits_{i=1}^3 (1+\kappa_i) \left(\frac{r_+}{r}\right)^4, 
\ee
thus $\sx=\sr=2$, $\sgh=4$, and $m = r_+^4 \prod_{i=1}^3 (1+\kappa_i) \propto \epsilon_0$ from \eq{energy_density}. The fall-off behavior satisfies the condition~(\ref{eq:fall-off_index}). Note that $m$ becomes complicated if one wants to write it in terms of charges and temperature. From \eq{def-index}, we get
\be
\nu_{\rm R}=\frac{\sx+\sr-2}{2\sgh}=\frac{1}{4}~,
\ee
and \eq{L_max-behavior} is written as
\be
\ls{R} \propto \frac{\sqrt{N_c}}{(\epsilon_0\che)^{1/4}}~.
\ee
Moreover, note that the screening length is exactly the same as the \SAdS case {\it at a given energy density} because the expressions (\ref{eq:L_sin-approx})-(\ref{eq:def-index}) do not change. 
In particular, the results for ${\rm SAdS}_d$, (\ref{eq:perpendicular_result})
and (\ref{eq:parallel_result}), also hold for any R-charged black hole without
modifications.
This is because only the leading behavior of the metric matters by taking the
ultra-relativistic limit as we saw in \sect{formula}. The leading behavior
depends only on the energy density, not on the charge. [For example, see
\eq{RNAdS}.]
This suggest that it is more appropriate to define the screening length $L_s$ as
\be
L_s \propto \frac{f(v)}{\epsilon_0^{1/4}} (1-v^2)^{1/4}
\ee
rather than using the temperature for generic case. 

\subsection{The other dimensions}

Let us briefly look at gauge/gravity duals in the other dimensions to see the dimensional dependence on the screening length. 
As an example, consider the \SAdSd black hole given by
\be
  ds^2 = - \left(\frac{r}{R}\right)^2 
\left\{ 1-\left(\frac{r_0}{r}\right)^{d} \right\} dt^2 
+ \frac{dr^2}{\left(\frac{r}{R}\right)^2 \{ 1-(\frac{r_0}{r})^{d} \} }  
+ \left(\frac{r}{R}\right)^2 (dx_1^2 + \cdots + dx_{d-1}^2)~.
\label{eq:SAdSd}
\ee
We assume $d\geq 4$. This black hole is dual to a $d$-dimensional conformal
field theory at finite temperature. But we do not specify the precise duals
since the main purpose here is just to look at dimensional dependence on the
screening length.%
\footnote{
Some examples are ${\rm SAdS}_4 \times S^7$ and ${\rm SAdS}_7 \times S^4$ in
M-theory, which correspond to the M2-brane and M5-brane, respectively. Of
course, it does not really make sense to use the fundamental string in the
context of M-theory. 
}

The temperature $T$ and the energy density $\epsilon_0$ of the black hole are given by 
\bea
T &=& \frac{d}{4\pi R^2} r_0~, \\
\epsilon_0 &=& \frac{d-1}{16\pi G_{d+1}}\frac{r_0^{d}}{R^{d+1}}~,
\eea
where $G_{d+1}$ is the $(d+1)$-dimensional Newton constant.

The fall-off behavior of the \SAdSd is
\be
g_{xx} \sim \left(\frac{r}{R}\right)^2~, \quad
g_{rr} \sim \left(\frac{r}{R}\right)^{-2}~, \quad
h \sim \left(\frac{r_0}{r}\right)^d~,
\ee
thus $\sx=\sr=2$, $\sgh=d$, and $m = r_0^d \propto \epsilon_0$. The fall-off behavior satisfies the condition~(\ref{eq:fall-off_index}). Thus, $\nu_{\rm SAdS}=1/d$ and
\be
L_s \propto \frac{1}{T (\che)^{1/d}}
\propto R\left(\frac{1}{G_{d+1} R}\right)^{1/d} \frac{1}{(\epsilon_0\che)^{1/d}}~.
\ee

\section{Non-conformal theories}\label{sec:nonconf}

We now move on to non-conformal theories; We will see in the following
examples that the exponent $\nu$  deviates from $1/d$.

\subsection{Klebanov-Tseytlin geometry}\label{sec:kt}

As an example of non-conformal theories, consider the KT geometry, which is dual
to ${\cal N}=1$ cascading $SU(K_*) \times SU(K_*+P)$ gauge theory. The finite
temperature solution,  for temperatures high above the
deconfining transition, was constructed  in Refs.~\cite{Gubser:2001ri,Buchel:2001gw,Buchel:2000ch}. In this regime, the theory is parametrized by the deformation parameter $\deltakt$:
\be
\deltakt := \frac{P^2}{\Ks} \ll 1~.
\ee
The 10-dimensional metric (in
the Einstein metric) is given by
\be
ds^2_{\rm E} = \frac{\sqrt{8a/\Ks}}{\sqrt{z}} e^{2P^2\eta} 
\left\{ - (1-z) dt^2 + dx_i^2 \right\}
+ \frac{\sqrt{\Ks}}{32} e^{-2P^2(\eta-5\xi)} \frac{dz^2}{z^2(1-z)} + \cdots, 
\ee
where $\cdots$ stands for the compact five-dimensional part and the radial coordinate $z$ runs from the horizon $z=1$ to the asymptotic infinity $z\rightarrow 0$. The solution is valid to the first order in $P^2/\Ks \ll 1$. The solution is known for all range of $z$ ($0<z\leq1$). In Ref.~\cite{PandoZayas:2006sa}, it was argued that to study in detail the solution in the interval  $z < z_c$, where $z_c$ is a very small but nonzero number,  a numerical analysis is necessary.  We will only be concerned with the leading terms in the metric which are  summarized in  Eqs.~(5.22), (5.30), and (5.31) of Ref.~\cite{Gubser:2001ri}:
\bea
\eta &\sim& \frac{\log z-1}{8\Ks}+O(z)~, \\
\xi &\sim& O(z)~, \\
\phi &\sim& O(z\,\ln z)~.
\eea
As discussed at the end of \sect{formula}, one can use the coordinate $z$ to find the exponent. Then,
\be
\sx =\frac{P^2}{4\Ks}-\frac{1}{2}~, \qquad
\sigma_z =\frac{P^2}{4\Ks}+2~, \qquad
\sgh =-1~.
\ee 
Thus, 
\be
\nu_{\rm KT} = \frac{1-\deltakt}{4}<\frac{1}{4}~.
\label{eq:kt_exponent}
\ee
Note that even though KT geometry is dual to a four dimensional gauge theory,  the
exponent we find in this case deviates from  $1/4$. This deviation
is measured by $ \deltakt $ which is a non-conformality parameter. Thus,
the fact that the scaling exponent is less than  $1/d$ for a KT background is
intimately related with the non-conformal nature of the theory.  

\subsection{D$p$-branes}

In the previous subsection we saw that the exponent deviates from $1/d$ for a
non-conformal theory. In order to see how large the deviation can be, it is
desirable to study theories with strong deviation from the conformality. When
the deviation is small, many examples are known. Unfortunately, few theories are
known when the deviation is large;  the D$p$-brane is one such  example.
The D$p$-brane background  is dual to the $(p+1)$-dimensional SYM with 16
supercharges. (For a recent discussion of this duality, see
Ref.~\cite{Maeda:2005cr} and references therein.) In the string metric, the
near-horizon limit of the D$p$-brane geometry  (for $p<7$) is given by
\bea
ds^2 &=& - \left(\frac{r}{R}\right)^{\frac{7-p}{2}} 
\left\{ 1 - \left(\frac{r_0}{r} \right)^{7-p} \right\} dt^2 
+ \left(\frac{r}{R}\right)^{\frac{7-p}{2}} (dx_1^2 + \cdots + dx_p^2) 
\nonumber \\
&&+ \frac{dr^2}{(\frac{r}{R})^{\frac{7-p}{2}} 
\{ 1 - \left(\frac{r_0}{r}\right)^{7-p} \}} 
+ R^2 \left(\frac{r}{R}\right)^{\frac{p-3}{2}} d\Omega_{8-p}^2~.
\eea
The $p=3$ case is the \SAdS solution.
The temperature $T$ and the energy density $\epsilon_0$ of the black hole are given by
\bea
T &=& \frac{(7-p)}{4\pi}\frac{r_0^{\frac{5-p}{2}}}{R^{\frac{7-p}{2}}}~, \\
\epsilon_0 &=& \frac{9-p}{32\pi G_{10}} \Omega_{8-p} r_0^{7-p}
\propto \lambda^{\frac{p-3}{5-p}} N_c^2 T^{\frac{2(7-p)}{5-p}}~.
\eea

The fall-off behavior of the D$p$-branes is
\be
g_{xx} \sim \left(\frac{r}{R}\right)^\frac{7-p}{2}~, \quad
g_{rr} \sim \left(\frac{r}{R}\right)^{-\frac{7-p}{2}}~, \quad
h \sim \left(\frac{r_0}{r}\right)^{7-p}~,
\ee
thus $\sx=\sr=(7-p)/2$, $\sgh=7-p$, and $m = r_0^{7-p} \propto \epsilon_0$. The fall-off behavior satisfies the condition~(\ref{eq:fall-off_index}). Thus, 
\be
\nu_{Dp}=\frac{5-p}{2(7-p)}
\label{eq:dp_exponent}
\ee
and
\be
\ls{Dp} 
\propto \frac{1}{T(\che)^{\frac{5-p}{2(7-p)}}} 
\propto \left( \frac{\lambda^{\frac{p-3}{5-p}}N_c^2}{\epsilon_0\che} \right)^{\frac{5-p}{2(7-p)}}~.
\label{eq:dp_screen}
\ee
For $\theta=\pi/2$, it is easy to carry out the integral (\ref{eq:L_sin-approx}), and the maximum value of $L$ is given by
\bea
\lefteqn{\ls{Dp}(\theta=\pi/2) =}
\nonumber \\
&&\mbox{}\frac{2\sqrt{\pi}}{\sqrt{(5-p)(7-p)}}
\left(\frac{5-p}{2(6-p)}\right)^{\frac{6-p}{7-p}}
\frac{\Gamma(1-\frac{1}{7-p})}{\Gamma(\frac{3}{2}-\frac{1}{7-p})} 
\frac{R}{\{ (\frac{r_0}{R})^{7-p}\che \}^{\frac{5-p}{2(7-p)}}}~.
\eea
One can check that the $p=3$ case agrees with the results in \sect{sads}.
However, for $p\ne 3$, the screening length does not behave as
$\epsilon_0^{-1/(p+1)}$ as one can see from \eq{dp_exponent}. This is partly
related to the fact that the D$p$-brane sometimes ceases to be a
$(p+1)$-dimensional theory. For example, the D4-brane is a 6-dimensional theory
in disguise. In fact, setting $p=4$, one gets $L_s \sim
(\epsilon_0\cosh^2\eta)^{-1/6}$. This is precisely the result one would expect
for a 6-dimensional (conformal) theory. 

As is well-known, such a transition does occur since the type IIA description
becomes a bad description in the ultraviolet (as $r\rightarrow\infty$)
and the  M-theory description takes over. The M5-brane, which is conformal, 
becomes the natural object to consider. So, this behavior is intimately
related to the fact that the D$p$-brane is nonconformal.
The deviation from the conformal value is
\be
\nu_{Dp} - \frac{1}{p+1} = \frac{-(p-3)^2}{2(7-p)(p+1)} \leq 0~,
\ee
so the exponent is always smaller than the conformal value (except $p=3$).

\subsection{The scaling exponent versus the speed of sound}\label{sec:cs}

As we have seen, the scaling exponent $\nu$ deviates from the
conformal value and the deviation may be parametrized by the deformation
parameter. The speed of sound $c_s$ in a non-conformal gauge theory  plasma
also deviates from the conformal value $1/3$ and the deviation may be again
parametrized by the deformation parameter. Thus, it is reasonable to look for 
an expression of  $\nu$ in terms of $c_s$. We obtain such an expression in this
subsection. 

For the KT geometry, the speed of sound has been computed in Ref.~\cite{Buchel:2005cv}:
\be
c_s^2 = \frac{1}{3}\left(1-\frac{4}{3} \deltakt\right)
+O\left(\deltakt^2\right)~.
\ee
Using \eq{kt_exponent}, one gets
\be
4\nu = 1-\frac{3}{4}(1-3c_s^2)+\cdots~.
\label{eq:mu_vs_cs}
\ee

Interestingly, the above relation is valid not only for the KT geometry, but
also, in some sense, for the D$p$-brane solution. For  D$p$-branes, one can
regard 
\be
\deltadp:= p-3 
\ee
as the deformation parameter; In reality, $p$ is of course an integer and we do not know if the $\deltadp$-expansion can be justified, but let us proceed. The speed of sound for the D$p$-brane has never been computed for all values of $p$, so we make a simple use of the thermodynamic relation $p=T s -\epsilon$ (for zero chemical potential). The pressure obtained from the thermodynamic relation is given by
\be
p=\frac{5-p}{32\pi G_{10}} \Omega_{8-p} r_0^{7-p}~,
\qquad
s=\frac{1}{4G_{10}} \Omega_{8-p} R^{\frac{7-p}{2}} r_0^{\frac{9-p}{2}}~,
\ee
where we also list the entropy density for completeness. Then, the speed of sound $c_s^2 :=\partial p/\partial \epsilon$ is given by
\be
c_s^2 = \frac{5-p}{9-p} 
= \frac{1}{3}\left(1-\frac{\deltadp}{3}\right)
+O\left(\deltadp^2\right).
\ee
For $p=3$, $c_s^2=1/3$ as expected. For $p=4$, $c_s^2=1/5$, which 
coincides with the explicit computation in Ref.~\cite{Benincasa:2006ei}. 
The value $1/5$ rather than $1/4$ is due to the underlying higher-dimensional 
nature of the D4-brane. For $p=1$, $c_s^2=1/2$ because of the same reason 
(The D1-brane is the conformal M2-brane in disguise). On the other hand, 
$p=2$ gives $c_s^2=3/7$, which does not seem to have such a simple interpretation.

Using the speed of sound, one gets \eq{mu_vs_cs} for the D$p$-brane as well. In fact, for this system, a simple thermodynamic argument implies the relation \cite{Natsuume:2007vc}. Thus, we conjecture that
\eq{mu_vs_cs} is true for theories with one deformation parameter in general. In previous subsections, we
found the exponent $\nu$ becomes smaller than the conformal value. From the
point of view of \eq{mu_vs_cs}, this is because the speed of sound often decreases from $1/3$ for nonconformal theories. 

\section{Discussion}\label{sec:discussion}

We found that the leading behavior of the screening length for conformal
theories  is given by $(\mbox{boosted energy density})^{-1/d}$. This behavior does not
survive for nonconformal theories. Thus, in principle, we would not expect that 
 $L_s\propto (\mbox{boosted energy density})^{-1/4 }$ will apply for  QCD.
However, lattice results indicate that the speed of sound becomes  close to
$1/3$ for $T \geq 2T_c$ (see Ref.~\cite{Karsch:2006sm} for a summary of results
by various groups). This implies that QCD may be approximately regarded as a
conformal theory for such a range of temperatures. And this is precisely  the
range of temperatures of current and near-future experiments. Therefore, the
conformal result may  still be useful for modelling charmonium suppression in
heavy ion collisions. 

Even if the scaling exponent for QCD turns out to deviate
from $1/4$, we expect the deviation to be  proportional to the non-conformality
parameter. It would be interesting to study  the deviation from conformality 
for QCD (theoretically and experimentally) and to find a gravity dual with a 
exponent similar to QCD. 
One can make a simple estimate of the deviation for QCD.%
\footnote{We thank Krishna Rajagopal for the suggestion on this point.}
According to the lattice results cited in Ref.~\cite{Karsch:2006sm}, all groups
 roughly predict $1/3-c_s^2 \sim 0.05$ around $2T_c$. Bearing in mind that our
 results are valid to large-$N_c$ theories and not to QCD, \eq{mu_vs_cs} 
gives $\nu \sim 0.22$. It would be interesting to compare this number with QCD 
calculations and experimental results.

We also found that the exponent becomes smaller than $1/d$ for nonconformal
theories. It would be interesting to check if this is also true for  other
nonconformal theories. One way to understand this phenomenon is to use the
relation between the exponent and the speed of sound (\ref{eq:mu_vs_cs}). The
exponent becomes smaller since the speed of sound often decreases 
from $1/3$ for nonconformal theories. It would be interesting to check that
 (\ref{eq:mu_vs_cs}) 
 also holds for  other nonconformal theories. 

Unlike QCD, none of the backgrounds studied here  include dynamical quarks.
Until recently, only solutions with   
$N_f/N_c \rightarrow 0$ were known. Lately, there has been an effort to find
backgrounds that will include dynamical quarks beyond this approximation
\cite{Casero:2006pt,Kirsch:2005uy,Burrington:2005zd}. It  would
be interesting to explore the behavior of the screening length in these models. 
Ideally, one would also like to explore  the screening length in non 
supersymmetric plasmas.~\footnote{In \cite{Peeters:2006iu}
 the authors studied the dissociation of  large spin mesons in  a 
confining non-supersymmetric model.}
 
For \SAdS and R-charged black holes, the screening length in the  ultrarelativistic limit is   the same at a given energy density. As we saw in \sect{formula}, only the leading behavior of the metric matters in the ultrarelativisitc limit. Therefore the leading behavior depends only on the energy density, not on the charge.

The screening length at finite chemical potential is the same as the one at zero chemical potential, 
but one should keep in mind that this is valid only in the ultrarelativistic limit. They are certainly not the same for generic $v$.
For arbitrary $v$, $L_s$ is expected to be  
\be
L_s \propto \frac{f(v)}{\epsilon_0^{1/4}} (1-v^2)^{1/4}~,
\ee
where $f(v)$ is a slowly varying function of order one (by defining the function appropriately). 
A  numerical computation is necessary to determine $f(v)$. We have carried out such an analysis as well
and  will present a  detailed discussion of the  numerical results elsewhere,
but there are three things to be noted. 
\begin{itemize}
\item The function $f(v)_R$ is order one and $f(v)_R \sim f(v)_{R=0}$, where $f(v)_R$ and $f(v)_{R=0}$ are the ones for finite chemical potential and for zero chemical potential, respectively.
\item The $f(v)_R$-curve is approaching to $f(v)_{R=0}$ in the ultrarelativistic limit as we saw in this paper.
\item The chemical potential dependence is very mild.
\end{itemize}

We hope that our analysis in the ultrarelativistic limit will be useful to understand the screening length behavior  for different gauge theories.

\acknowledgments
It is a pleasure to thank Alberto G\"uijosa, Testuo Hatsuda, Tetsufumi Hirano, Kazunori Itakura, Kengo Maeda, Tetsuo Matsui, Osamu Morimatsu, and Berndt M\"{u}ller for useful conversations. Elena C\'aceres thanks the
Theory Group at the University of Texas at Austin for hospitality
during the completion of this work. Her research is supported in
part by the National Science Foundation under Grant Nos.
PHY-0071512 and PHY-0455649, by  Ram\'on Alvarez Bulla grant \# 447/06  and CONACyT grant \# 50760. The research of M.N.\ was supported in part by the Grant-in-Aid for Scientific Research (13135224) from the Ministry of Education, Culture, Sports, Science and Technology, Japan.

\vspace*{0.3cm}
{\bf Note added}:
After this work was completed, we were informed that Liu, Rajagopal and
Wiedemann have also carried out the ultrarelativistic analysis for \SAdS  described in \sect{sads}.
We would like to thank them for communicating their unpublished results and for discussion.


\footnotesize

\begin{thebibliography}{99}


\bibitem{Teaney:2003pb}
D.~Teaney,
``Effect of shear viscosity on spectra, elliptic flow, and Hanbury Brown-Twiss
radii,''
Phys.\ Rev.\ C {\bf 68}, 034913 (2003).

\bibitem{Hirano:2005wx}
T.~Hirano and M.~Gyulassy,
``Perfect fluidity of the quark gluon plasma core as seen through its
dissipative hadronic corona,''
arXiv:nucl-th/0506049.

\bibitem{Policastro:2001yc}
  G.~Policastro, D.~T.~Son and A.~O.~Starinets,
  ``The shear viscosity of strongly coupled N = 4 supersymmetric Yang-Mills
  plasma,''
  Phys.\ Rev.\ Lett.\  {\bf 87}, 081601 (2001)
  [arXiv:hep-th/0104066].
  

\bibitem{Kovtun:2003wp}
  P.~Kovtun, D.~T.~Son and A.~O.~Starinets,
  ``Holography and hydrodynamics: Diffusion on stretched horizons,''
  JHEP {\bf 0310}, 064 (2003)
  [arXiv:hep-th/0309213].

\bibitem{Buchel:2003tz}
  A.~Buchel and J.~T.~Liu,
  ``Universality of the shear viscosity in supergravity,''
  Phys.\ Rev.\ Lett.\  {\bf 93}, 090602 (2004)
  [arXiv:hep-th/0311175].

\bibitem{Kovtun:2004de}
  P.~Kovtun, D.~T.~Son and A.~O.~Starinets,
  ``Viscosity in strongly interacting quantum field theories from black hole
  physics,''
  Phys.\ Rev.\ Lett.\  {\bf 94}, 111601 (2005)
  [arXiv:hep-th/0405231].

\bibitem{Buchel:2004qq}
  A.~Buchel,
  ``On universality of stress-energy tensor correlation functions in
  supergravity,''
  Phys.\ Lett.\ B {\bf 609}, 392 (2005)
  [arXiv:hep-th/0408095].
  
  
\bibitem{Mas:2006dy}
  J.~Mas,
  ``Shear viscosity from R-charged AdS black holes,''
  JHEP {\bf 0603} (2006) 016
  [arXiv:hep-th/0601144].

\bibitem{Son:2006em}
  D.~T.~Son and A.~O.~Starinets,
  ``Hydrodynamics of R-charged black holes,''
  JHEP {\bf 0603} (2006) 052
  [arXiv:hep-th/0601157].

\bibitem{Saremi:2006ep}
  O.~Saremi,
  ``The viscosity bound conjecture and hydrodynamics of M2-brane theory at
  finite chemical potential,''
  arXiv:hep-th/0601159.

\bibitem{Maeda:2006by}
  K.~Maeda, M.~Natsuume and T.~Okamura,
  ``Viscosity of gauge theory plasma with a chemical potential from AdS/CFT correspondence,''
  Phys.\ Rev.\ D {\bf 73} (2006) 066013
  [arXiv:hep-th/0602010].

\bibitem{Natsuume:2007qq}
  M.~Natsuume,
  ``String theory and quark-gluon plasma,''
  arXiv:hep-ph/0701201.
  

\bibitem{Liu:2006ug}
  H.~Liu, K.~Rajagopal and U.~A.~Wiedemann,
  ``Calculating the jet quenching parameter from AdS/CFT,''
  arXiv:hep-ph/0605178.

\bibitem{Herzog:2006gh}
  C.~P.~Herzog, A.~Karch, P.~Kovtun, C.~Kozcaz and L.~G.~Yaffe,
  ``Energy loss of a heavy quark moving through N = 4 supersymmetric Yang-Mills
  plasma,''
  arXiv:hep-th/0605158.

\bibitem{Casalderrey-Solana:2006rq}
  J.~Casalderrey-Solana and D.~Teaney,
  ``Heavy quark diffusion in strongly coupled N = 4 Yang Mills,''
  arXiv:hep-ph/0605199.

\bibitem{Gubser:2006bz}
  S.~S.~Gubser,
  ``Drag force in AdS/CFT,''
  arXiv:hep-th/0605182.

\bibitem{Buchel:2006bv}
  A.~Buchel,
  ``On jet quenching parameters in strongly coupled non-conformal gauge
  theories,''
  arXiv:hep-th/0605178.

\bibitem{Herzog:2006se}
  C.~P.~Herzog,
  ``Energy Loss of Heavy Quarks from Asymptotically AdS Geometries,''
  arXiv:hep-th/0605191.

\bibitem{Caceres:2006dj}
  E.~Caceres and A.~Guijosa,
  ``Drag Force in Charged N=4 SYM Plasma,''
  arXiv:hep-th/0605235.

\bibitem{Vazquez-Poritz:2006ba}
  J.~F.~Vazquez-Poritz,
  ``Enhancing the jet quenching parameter from marginal deformations,''
  arXiv:hep-th/0605296.

\bibitem{Caceres:2006as}
  E.~Caceres and A.~Guijosa,
  ``On drag forces and jet quenching in strongly coupled plasmas,''
  arXiv:hep-th/0606134.

\bibitem{Lin:2006au}
  F.~L.~Lin and T.~Matsuo,
  ``Jet quenching parameter in medium with chemical potential from AdS/CFT,''
  arXiv:hep-th/0606136.

\bibitem{Avramis:2006ip}
  S.~D.~Avramis and K.~Sfetsos,
  ``Supergravity and the jet quenching parameter in the presence of R-charge
  densities,''
  arXiv:hep-th/0606190.

\bibitem{Armesto:2006zv}
  N.~Armesto, J.~D.~Edelstein and J.~Mas,
  ``Jet quenching at finite `t Hooft coupling and chemical potential from
  AdS/CFT,''
  arXiv:hep-ph/0606245.

\bibitem{Sin:2004yx}
  S.~J.~Sin and I.~Zahed,
  ``Holography of radiation and jet quenching,''
  Phys.\ Lett.\ B {\bf 608} (2005) 265
  [arXiv:hep-th/0407215].
  
\bibitem{Liu:2006nn}
  H.~Liu, K.~Rajagopal and U.~A.~Wiedemann,
  ``An AdS/CFT calculation of screening in a hot wind,''
  arXiv:hep-ph/0607062.

\bibitem{Chernicoff:2006hi}
  M.~Chernicoff, J.~A.~Garcia and A.~Guijosa,
  ``The Energy of a Moving Quark-Antiquark Pair in an N=4 SYM Plasma,''
  arXiv:hep-th/0607089.

\bibitem{Chu:1988wh}
  M.~C.~Chu and T.~Matsui,
   ``Dynamic Debye Screening for a heavy quark-antiquark pair traversing a quark-gluon plasma,''
  Phys.\ Rev.\ D {\bf 39} (1989) 1892.
  

\bibitem{Rey:1998bq}
  S.~J.~Rey, S.~Theisen and J.~T.~Yee,
  ``Wilson-Polyakov loop at finite temperature in large N gauge theory and
  anti-de Sitter supergravity,''
  Nucl.\ Phys.\ B {\bf 527} (1998) 171
  [arXiv:hep-th/9803135].

\bibitem{Brandhuber:1998bs}
  A.~Brandhuber, N.~Itzhaki, J.~Sonnenschein and S.~Yankielowicz,
  ``Wilson loops in the large N limit at finite temperature,''
  Phys.\ Lett.\ B {\bf 434} (1998) 36
  [arXiv:hep-th/9803137].

\bibitem{Gubser:1996de}
  S.~S.~Gubser, I.~R.~Klebanov and A.~W.~Peet,
  ``Entropy and Temperature of Black 3-Branes,''
  Phys.\ Rev.\ D {\bf 54} (1996) 3915
  [arXiv:hep-th/9602135].
  

\bibitem{Gubser:2001ri}
  S.~S.~Gubser, C.~P.~Herzog, I.~R.~Klebanov and A.~A.~Tseytlin,
  ``Restoration of chiral symmetry: A supergravity perspective,''
  JHEP {\bf 0105} (2001) 028
  [arXiv:hep-th/0102172].

\bibitem{Buchel:2001gw}
  A.~Buchel, C.~P.~Herzog, I.~R.~Klebanov, L.~A.~Pando Zayas and A.~A.~Tseytlin,
  ``Non-extremal gravity duals for fractional D3-branes on the conifold,''
  JHEP {\bf 0104} (2001) 033
  [arXiv:hep-th/0102105].


\bibitem{Buchel:2000ch}
  A.~Buchel,
  ``Finite temperature resolution of the Klebanov-Tseytlin singularity,''
  Nucl.\ Phys.\ B {\bf 600} (2001) 219
  [arXiv:hep-th/0011146].

\bibitem{PandoZayas:2006sa}
  L.~A.~Pando Zayas and C.~A.~Terrero-Escalante,
  ``Black holes with varying flux: A numerical approach,''
  arXiv:hep-th/0605170.

\bibitem{Maeda:2005cr}
K.~Maeda, M.~Natsuume and T.~Okamura,
``Quasinormal modes for nonextreme Dp-branes and thermalizations of
super-Yang-Mills theories,''
Phys.\ Rev.\ D {\bf 72}, 086012 (2005)
[arXiv:hep-th/0509079].

\bibitem{Buchel:2005cv}
  A.~Buchel,
  ``Transport properties of cascading gauge theories,''
  Phys.\ Rev.\ D {\bf 72} (2005) 106002
  [arXiv:hep-th/0509083].

\bibitem{Benincasa:2006ei}
  P.~Benincasa and A.~Buchel,
  ``Hydrodynamics of Sakai-Sugimoto model in the quenched approximation,''
  arXiv:hep-th/0605076.
  
\bibitem{Natsuume:2007vc}
  M.~Natsuume and T.~Okamura,
  ``Screening length and the direction of plasma winds,''
  arXiv:0706.0086 [hep-th].


\bibitem{Karsch:2006sm}
  F.~Karsch,
  ``Lattice QCD at high temperature and the QGP,''
  arXiv:hep-lat/0601013.
  
\bibitem{Casero:2006pt}
  R.~Casero, C.~Nunez and A.~Paredes,
  ``Towards the string dual of N = 1 SQCD-like theories,''
  Phys.\ Rev.\ D {\bf 73}, 086005 (2006)
  [arXiv:hep-th/0602027].
  
\bibitem{Kirsch:2005uy}
  I.~Kirsch and D.~Vaman,
  ``The D3/D7 background and flavor dependence of Regge trajectories,''
  Phys.\ Rev.\ D {\bf 72}, 026007 (2005).
  
\bibitem{Burrington:2005zd}
  B.~A.~Burrington, J.~T.~Liu, M.~Mahato and L.~A.~Pando Zayas,
   ``Towards supergravity duals of chiral symmetry breaking in Sasaki-Einstein
  cascading quiver theories,''
  JHEP {\bf 0507} (2005) 019
  [arXiv:hep-th/0504155].

\bibitem{Peeters:2006iu}
  K.~Peeters, J.~Sonnenschein and M.~Zamaklar,
  ``Holographic melting and related properties of mesons in a quark gluon
  plasma,''
  arXiv:hep-th/0606195.

\end{thebibliography}
\end{document}